%% file: main_revised.tex
\documentclass[%
 reprint,
superscriptaddress,
 amsmath,amssymb,
 aps,
]{revtex4-2}
\usepackage{physics}
\usepackage{todonotes}
\usepackage{graphicx}
\usepackage{dcolumn}
\usepackage{bm}
\usepackage{xcolor}

\newcommand{\be}{\begin{equation}}
\newcommand{\ee}{\end{equation}}
\newcommand{\bea}{\begin{eqnarray}}
\newcommand{\eea}{\end{eqnarray}}
\newcommand{\ciss}{{CISS}} 
\newcommand{\bM}{{\mathbf M}} 
\newcommand{\soc}{{SOC}} 
\newcommand{\ci}{{\mathfrak i}}

\begin{document}

\preprint{APS/123-QED}

\title{Theory of Chirality Induced Spin Selectivity: Progress and Challenges}

\author{Ferdinand Evers}
\email{ferdinand.evers@physik.uni-regensburg.de}
 \affiliation{Institute of Theoretical Physics, University of Regensburg, 93040 Regensburg, Germany}
 
\author{Amnon Aharony}
\affiliation{School of Physics and Astronomy, Tel Aviv University, Tel Aviv 6997801, Israel}

\author{Nir Bar-Gill}
\affiliation{Department of Applied Physics, Racah Institute of Physics, The Hebrew University of Jerusalem, Jerusalem 9190401, Israel}

\author{Ora Entin-Wohlman}
\affiliation{Raymond and Beverly Sackler School of Physics and Astronomy, Tel Aviv University, Tel Aviv 6997801, Israel}

\author{Per Hedegård}
\affiliation{Niels Bohr Institute, University of Copenhagen, DK-2100 Copenhagen, Denmark}
 
 \author{Oded Hod}
 \affiliation{Department of Physical Chemistry, School of Chemistry, The Raymond and Beverly Sackler Faculty of Exact Sciences and The Sackler Center for Computational Molecular and Materials Science, Tel Aviv University, Tel Aviv 6997801, Israel}
 
 \author{Pavel Jelinek}
 \affiliation{Nanosurf Lab, Institute of Physics of the Czech Academy of Sciences, Prague 6, CZ 162 00, Czech Republic}
 
 \author{Grzegorz Kamieniarz}
 \affiliation{Department of Physics, Adam Mickiewicz University, Poznań 61-614, Poland}

 \author{Mikhail Lemeshko}
 \affiliation{IST Austria (Institute of Science and Technology Austria), Am Campus 1, 3400 Klosterneuburg, Austria}
 
 \author{Karen Michaeli}
 \affiliation{Department of Condensed Matter Physics, Weizmann Institute of Science, Rehovoth 7610001, Israel}
 
 \author{Vladimiro Mujica}
 \affiliation{School of Molecular Sciences, Arizona State University, Tempe, AZ 85287-1604, USA}
 
 \author{Ron Naaman} 
 \affiliation{Department of Chemical and Biological Physics, Weizmann Institute of Science, Rehovoth 76100, Israel}
 
 \author{Yossi Paltiel}
 \affiliation{Department of Applied Physics, Racah Institute of Physics, The Hebrew University of Jerusalem, Jerusalem 9190401, Israel}

 \author{Sivan Refaely-Abramson}
 \affiliation{Department of Molecular Chemistry and Materials Science, Weizmann Institute of Science, Rehovoth 76100, Israel}
 
 \author{Oren Tal}
 \affiliation{Department of Chemical and Biological Physics, Weizmann Institute of Science, Rehovoth 76100, Israel}
 
 \author{Jos Thijssen}
 \affiliation{3 Kavli Institute of Nanoscience Delft, Delft University of Technology, Lorentzweg 1, Delft, 2628 CJ The Netherlands}
 
 \author{Michael Thoss}
 \affiliation{Institute of Physics, University of Freiburg, Hermann-Herder-Str. 3, 79104 Freiburg, Germany}
 
 \author{Jan M. van Ruitenbeek}
 \affiliation{Huygens-Kamerlingh Onnes Laboratory, Leiden University, Niels Bohrweg 2, 2333 CA Leiden, Netherlands}
 
 \author{Latha Venkataraman}
 \affiliation{Department of Applied Physics and Department of Chemistry, Columbia University, New York, New York 10027, USA}
 
 \author{David H. Waldeck}
 \affiliation{Department of Chemistry, University of Pittsburgh, Pittsburgh, PA 15260, USA}
 
 \author{Binghai Yan}
 \affiliation{Department of Condensed Matter Physics, Weizmann Institute of Science, Rehovoth 7610001, Israel}
 
 \author{Leeor Kronik}
 \email{leeor.kronik@weizmann.ac.il}
 \affiliation{Department of Molecular Chemistry and Materials Science, Weizmann Institute of Science, Rehovoth 76100, Israel}

\begin{abstract}
We provide a critical overview of the theory of the chirality-induced spin selectivity (CISS) effect, i.e., phenomena in which the chirality of molecular species imparts significant spin selectivity to various electron processes. Based on discussions in a recently held workshop, and further work published since, we review the status of CISS effects - in electron transmission, electron transport, and chemical reactions. For each, we provide a detailed discussion of the state-of-the-art in theoretical understanding and identify remaining challenges and research opportunities. 
\end{abstract}

\maketitle

\input{Introduction.tex}
\input{Experiment.tex}

\input{Theory.tex}

\input{Conclusions_Outlook}


\bibliography{reviewRMP}
\bibliographystyle{bst/apsrev4-2_new}
\end{document}

%% file: Introduction.tex
 Chirality-induced spin selectivity (CISS), first discovered some two decades ago in the context of photoemission~\cite{Ray1999}, is now an umbrella term used to interpret a wide range of experimental phenomena in which the chirality of molecular species imparts significant spin selectivity to various electron processes~\cite{Naaman2015,Michaeli2016,Fontanesi2018,Naaman2019,Pop2019,Naaman2020,Waldeck2021,Yang2021}. The interplay between molecular handedness and electron spin suggests spin-orbit coupling (SOC) as a natural mechanism for CISS, given that it is easy to show that  ``current through a coil'' arguments yield an effect that is lower by many orders of magnitude. Early theoretical efforts have indeed confirmed that SOC may  provide a qualitative explanation for some aspects of the experimental findings. Quantitatively, however, such calculations have consistently predicted effects that were smaller by up to several orders of magnitude than those observed experimentally. While additional theoretical research efforts, described in more detail below, have shed more light on CISS, a complete quantitative theory of the effect remains elusive and its microscopic origins are insufficiently understood. 

The authors of this article have addressed this puzzle in a workshop hosted by the Weizmann Institute of Science in early 2020. In an attempt to form a new community with a joint research agenda, the workshop brought together theoretical and computational physicists and chemists of diverse backgrounds in different model Hamiltonian and first principles approaches, and with an interest in various relevant phenomena, including magnetic materials, strongly-correlated systems, topological materials, quantum transport, and molecular electronics and spintronics. These theorists were joined by several experimentalists with an interest in CISS. This article provides an overview of CISS theory, based on discussions in the workshop and further work published since. It provides a brief overview of the status of CISS experiments, followed by a detailed critical discussion of the state-of-the-art in theoretical understanding of CISS. Finally, it identifies remaining challenges and research opportunities.

%% file: Experiment.tex
\section{Survey of Experimental CISS Studies}

We start our considerations by presenting a brief survey of the status of CISS experimental work. This survey is not meant to be comprehensive. Rather, it aims to provide sufficient context for a meaningful analysis of the advantages and disadvantages of various theoretical approaches to CISS. For additional aspects of CISS studies, the reader is referred to past review articles, and references therein~\cite{Naaman2015,Michaeli2016,Fontanesi2018,Naaman2019,Pop2019,Naaman2020,Waldeck2021,Yang2021}.  

Relations between chirality and magnetic phenomena have a long history. Magnetically induced optical activity in crystals, as well as natural optical activity in chiral crystals, have been known since the nineteenth century, leading Pasteur himself to search, unsuccessfully, for a link between the two~\cite{Nakanishi1994}. It was not until 1997, however, that this link was finally found in the form of magneto-chiral dichroism, i.e., a difference in the magnetic optical activity of the two enantiomers of a chiral medium~\cite{Rikken1997}. CISS, discovered only two years after that~\cite{Ray1999}, takes a significant step further by establishing a direct link between chirality and spin, even in the absence of an external magnetic field and/or circularly polarized illumination. Generally speaking, CISS effects that have been observed since the original discovery can be divided into three broad categories: The first - also historically - involves transmission of unbound electrons (typically photoexcited from an underlying substrate) through a chiral medium to vacuum. The second - where most work to date has been done - involves transport of bound electrons, between leads, through a chiral medium. The third - and relatively new - category concerns relations between electron spin and chemical reactions.

\subsection{CISS in electron transmission}
\begin{figure*}
\includegraphics[width=8cm]{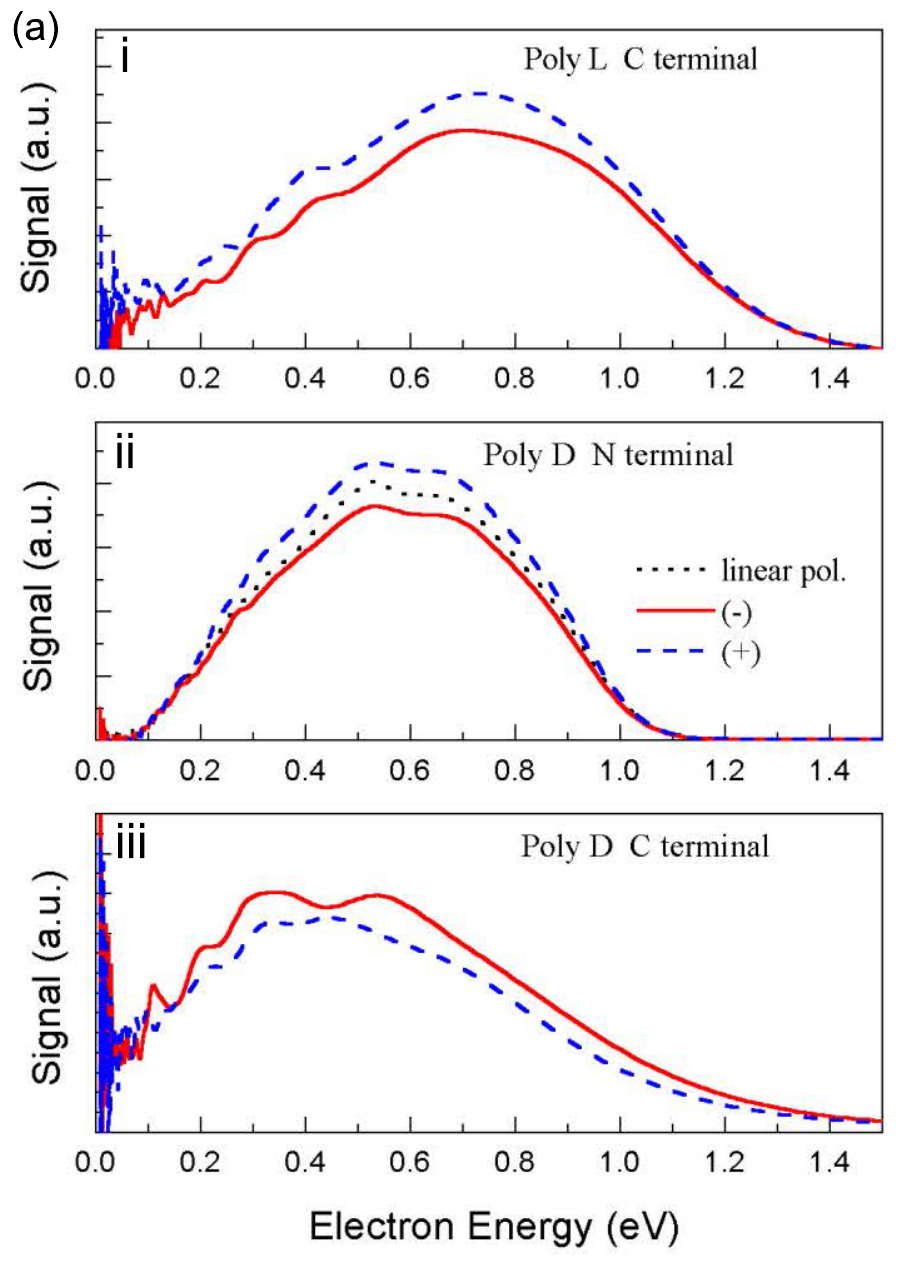}
\includegraphics[width=8cm]{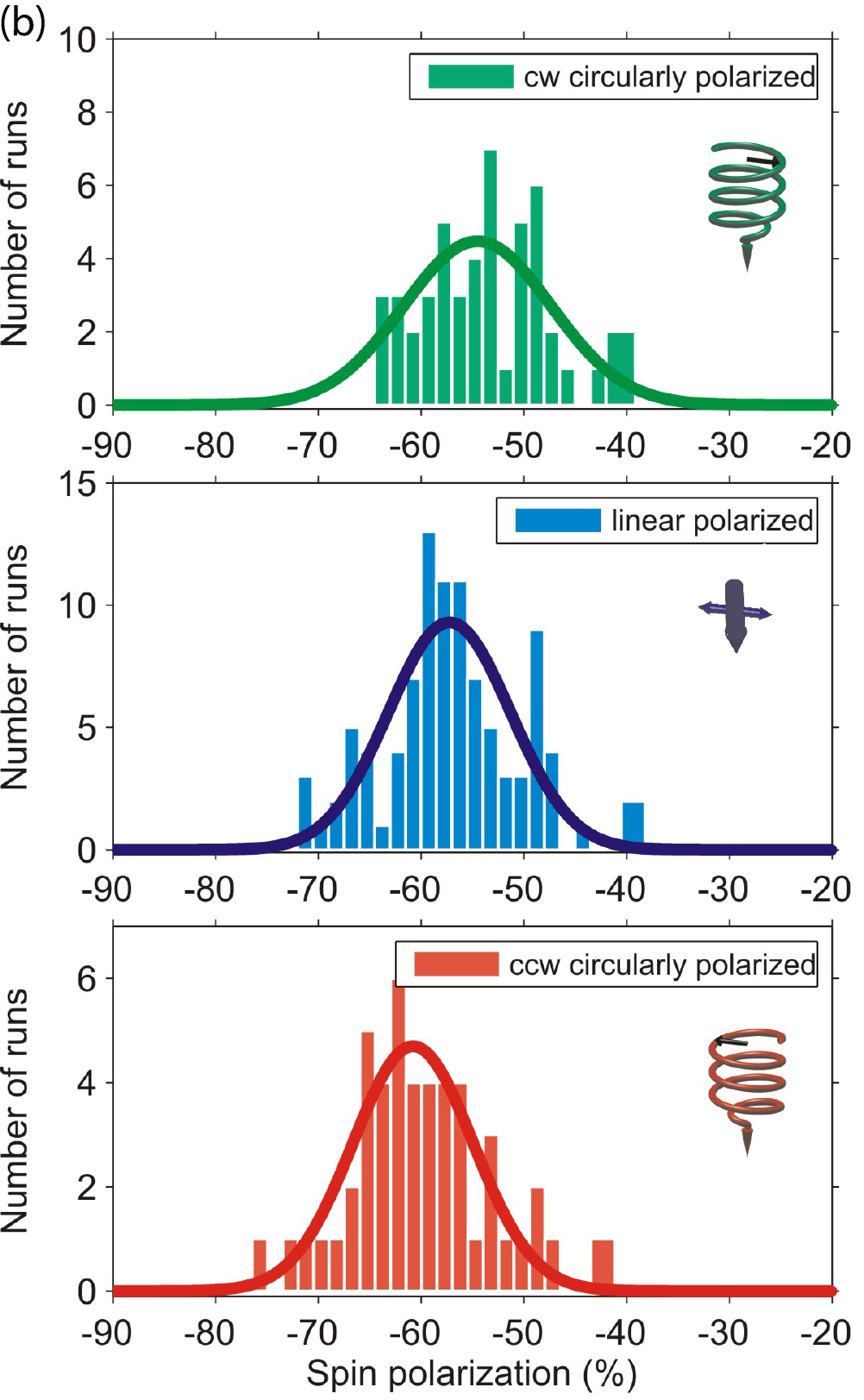}
\caption{(a) Energy distribution for photoelectrons transmitted through (i) L, C-terminus connected, (ii) D, N-terminus connected, and (iii) D, C-terminus connected        helical polyalanine films and excited using a cw (negative spin polarization; red, solid), ccw circularly (positive spin polarization; blue, dashed), or linearly (no spin polarization; black, dotted) polarized light. Taken from Ref.\ \protect\onlinecite{Carmeli2002}, used with permission. (b) Photoelectron polarization, measured for electrons ejected from a Au-coated substrate with a monolayer of 78-base pair double-stranded DNA, for cw circularly polarized light [(–54.5 $\pm$ 7.0\%); top, green], linearly polarized light [(-57.2 $\pm$ 5.9)\%; middle, blue] and ccw circularly polarized light [(-60.8 $\pm$ 5.8\%); bottom, red]. Taken from Ref.\ \protect\onlinecite{Goehler2011}, used with permission.}
\label{fig:transmission}
\end{figure*}

CISS in electron transmission was first observed experimentally by Ray {\it et al.}~\cite{Ray1999}, who considered how electrons, emitted from an Au substrate upon excitation with circularly polarized light, are transmitted through an organized adsorbed monolayer comprising chiral molecules. They found that electrons excited using clockwise (cw) light exhibited a significant asymmetry in the transmission probability as compared to those excited with counter-clockwise (ccw) light, with electrons excited using linearly polarized light exhibiting an intermediate transmission probability~\cite{Ray1999}. This is demonstrated here using the later work of Carmeli 
{\it et al.}~\cite{Carmeli2002}, shown in Fig.\ \ref{fig:transmission}a. In these studies, the illumination (cw or ccw) for which transmission through a polyalanine layer is preferred was found to depend on the handedness of the peptide (L or D, which exhibit left- and right-handed chirality, respectively). Based on the well-known connection between the direction (cw or ccw) of circularly polarized light and the spin polarization (up or down) of the excited electrons, it was inferred that the molecular chirality induces spin selectivity in the transmision, i.e., CISS is observed. This conjecture, while reasonable, took another 12 years to verify. This was finally achieved by G\"ohler {\it et al.}~\cite{Goehler2011}, who used a Mott polarimeter to measure directly the spin of photoelectrons transmitted from an Au substrate through a monolayer of double-stranded DNA. As shown in Fig.\ \ref{fig:transmission}b, significant spin-polarization was found even when the photoelectrons were generated with linearly polarized light. Further direct confirmation for the CISS effect in transmission came from the work of Ni\~no {\it et al.}~\cite{Nino2014}, who found significant differences in spin polarization between two enantiomers of the same molecule (1,2-diphenyl-1,2-ethanediol). 

Importantly, the essential effect does not depend on the specifics of the molecule and has been observed, e.g., in DNA of varying lengths, different oligopeptides, and helicenes (see Ref.\ \onlinecite{Nurenberg2019} and references therein). Interestingly, heptahelicene, composed of only carbon and hydrogen atoms and exhibiting only a single helical turn, was found to already show a longitudinal spin polarization of about 6\% to 8\% in the transmission of initially spin-balanced electrons~\cite{Kettner2018}. Note that, as also shown in Fig.\ \ref{fig:transmission}a for polyalanine, the direction of spin polarization can be inferred to depend not only on the handedness but also on the molecular dipole direction, as it changes sign depending on whether the molecule is bound to the substrate through the N-terminus (amine group side) or C-terminus (carboxyl group side)~\cite{Carmeli2002}.

\subsection{CISS in electron transport}
\begin{figure*}
\centering
\includegraphics[width=12cm]{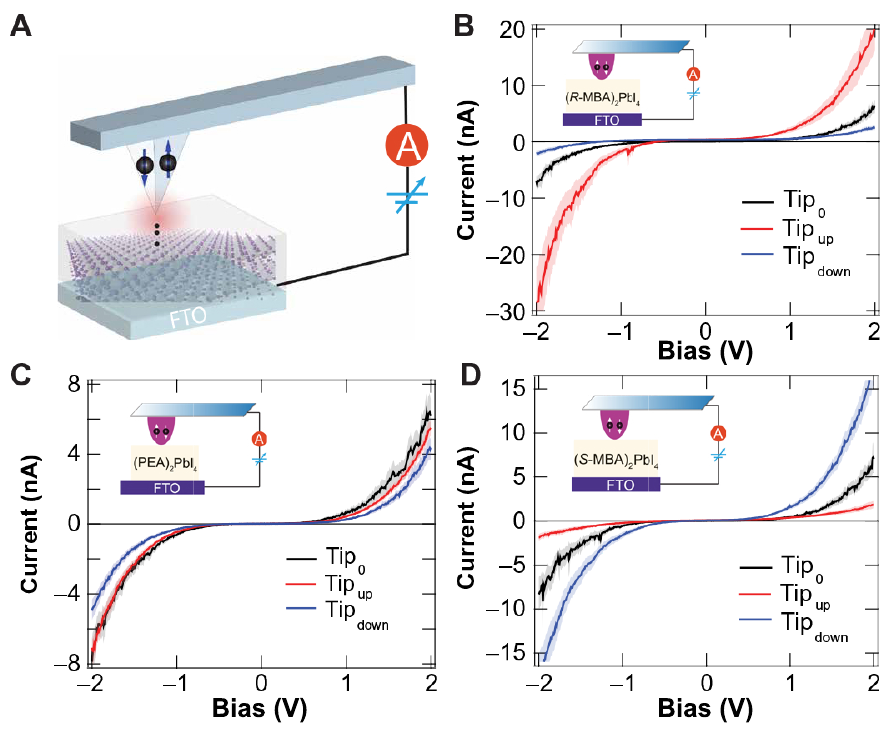}
\caption{(A) Setup for mCP-AFM room-temperature measurements of the chirality dependence in out-of-plane charge transport through chiral $\sim$50 nm two-dimensional hybrid perovskite (MBA)$_2$PbI$_4$ thin films deposited on fluorine-doped tin oxide (FTO) substrates. (B-D) Current-voltage curves for S (left-handed chiral), achiral, and R (right-handed chiral) films, with the tip magnetized north (blue), south (red), or not magnetized (black). The curves for each film were averaged over 100 scans and the shaded region around the lines marks the 95\% confidence limits for the average results. Taken from Ref.\ \protect\onlinecite{Lu2019}, used with permission.}
\label{fig:transport}
\end{figure*}

The above studies suggest that spin selectivity could be observed also for bound electrons traveling through a chiral medium, i.e., also for electron transport and not just for electron transmission. Experimental verification of this conjecture, however, requires proper contacting of the chiral medium - often an organic molecular layer - to metallic leads, which is not trivial. This was first accomplished by using a ferromagnetic substrate to inject spin (in this case to a DNA layer), with the tip of a conductive-probe atomic force microscope (CP-AFM) serving as the top contact~\cite{Xie2011}. As chiral media are often grown on non-magnetic substrates, more generally a magnetic CP-AFM (mCP-AFM) tip can be used, as shown in Fig.\ \ref{fig:transport}a~\cite{Lu2019}. Many other techniques have been used to detect spin imbalance due to CISS in electron transport. Two notable ones are magnetoresistance in a film where at least one electrode is ferromagnetic~\cite{Ravi2013,BenDor2014} and electrochemical measurements with a ferromagnetic electrode~\cite{Mishra2013}.

Several recent observations of CISS in electron transport are directly relevant for the theoretical discussion that follows. 
First, a sizable signal interpreted as being due to CISS effect has been reported in many articles~\cite{Naaman2015}. 
The effect has been observed in electron transport through many media, including different types of chiral molecular layers (including biological ones)~\cite{Xie2011,Abendroth2017,Mishra2019,Jia2020}, carbon nanotubes~\cite{Alam2015,Alam2017}, and chiral materials~\cite{Kulkarni2020,Lu2019,Lu2020,DiNuzzo2020,Mondal2020,Waldeck2021}. 
It has even been reported in single-molecule experiments using a break-junction~\cite{Aragones2017}.
Second, it has been repeatedly demonstrated that the effect generally increases with medium length~\cite{Mishra2020} and that it can be very large. For example, a spin-selectivity of up to 80\% was reported for electrons traversing ca.\ 2-6 $\mu$m-long self-assembled superhelical conducting polyaniline micro-fiber channels at room temperature~\cite{Jia2020}. Spin polarization exceeding 85\% was achieved using $\pi$-conjugated molecular materials based on coronene bisimide and tetra-amidated porphyrin cores appended with alkoxyphenyl groups~\cite{Kulkarni2020}. Even higher numbers were reported in recent studies of spin-dependent charge transport through 2D chiral hybrid organic-inorganic perovskite materials. Using ((R/S-)methylbenzylammonium (MBA) lead-iodide (see Fig.\ \ref{fig:transport}), a highest spin-polarization transport of up to 86\% was obtained~\cite{Lu2019} and with (R-/S)MBA-tin-iodide the efficiency was as high as 94\%~\cite{Lu2020}.
Third, as in CISS in transmission, also here the sign of the preferred spin depends on the direction of the molecular dipole~\cite{Eckshtain2016}, in addition to the usual dependence on the handedness.
Fourth, from a mechanistic point of view, two observations are important: CISS is repeatedly found to correlate with optical activity~\cite{Kulkarni2020,Mishra2020,Bloom2017} and CISS is prominent and clearly detected in the non-linear current-voltage regime (as also seen in Fig.\ \ref{fig:transport})~\cite{Kiran2017,Naaman2020,Yang2020}.

We also note that the spin-selectivity is increasingly explored practically for demonstrating spintronic effects and devices that can be used for logic and memory~\cite{Michaeli2017,Naaman2019}. Two recent examples are a spin filter~\cite{Suda2019} and magnetless Hall voltage measurements~\cite{Eckshtain2016,Mishra2019}.

Finally, it is important to note that a constructive and at the same time critical debate scrutinizing all measurements is still ongoing. In particular, CISS is not always observed in transport through chiral media and at present a broad consensus as to the precise experimental conditions under which a specific manifestation of CISS is expected does not exist.

\subsection{CISS in chemical reactions}
\begin{figure*}
\centering
\includegraphics[width=12 cm]{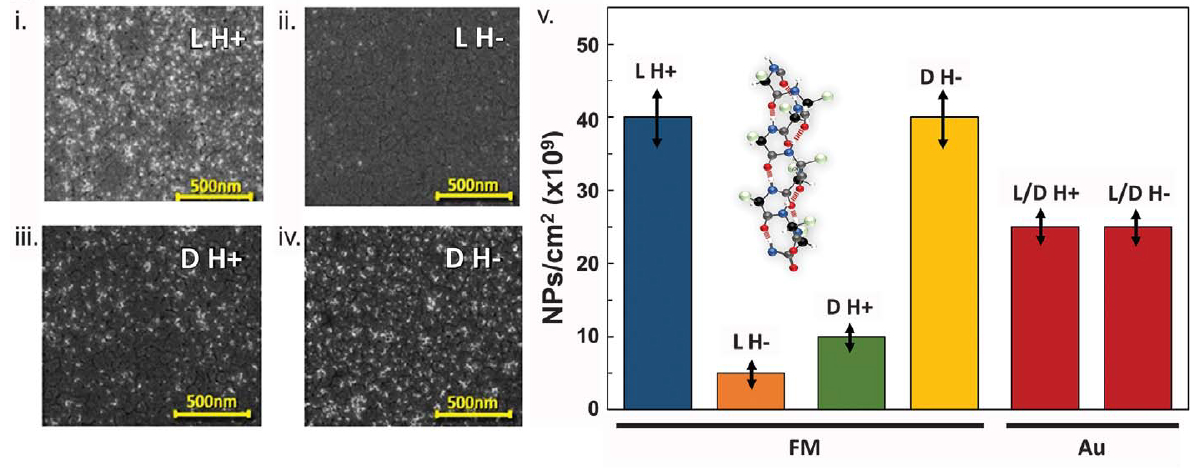}
\caption{Adsorption of a polyalanine oligopeptide [shown in inset of (v)] on ferromagnetic samples (silicon with a 1.8-nm Co film and a 5-nm Au film), magnetized with the
magnetic dipole pointing up (H+) or down (H–) relative to the substrate surface. SiO2 nanoparticles were attached to the adsorbed oligopeptides. Panels (i,ii) and (iii,iv) exhibit L-polyalanine and D-polyalanine, respectively, 
adsorbed for 2 s on a substrate magnetized up (i,iii) or down (ii,iv). Panel (v) summarizes the nanoparticle adsorption densities shown in (i) to (iv), compared with the adsorption density on Au with an applied
external magnetic field (red bars). Double-headed arrows represent error bars. The errors are the standard deviation among 10 measurements conducted on each of the 10 samples, hence a total of
100 measurements. Taken from Ref.\ \protect\onlinecite{Banerjee2018}, used with permission.}
\label{fig:enantio}
\end{figure*}

Beyond electron transmission or transport through a chemically stable medium, a third category of CISS uses the electron spin as an enantio-selective chemical reagent~\cite{Naaman2019b,Metzger2020}. The origins of this idea can be traced back to the work of Rosenberg {\it et al.}~\cite{Rosenberg2008}. They showed that use of low-energy spin-polarized secondary electrons, produced by irradiation of a magnetic substrate, results in different bond cleavage rates for R and S enantiomers of a chiral molecule adsorbed on the substrate. Based on CISS in transmission, the magnetic substrate was later replaced by a non-magnetic substrate with a chiral DNA overlayer acting as a spin filter, with similar consequences for enantio-selective chemistry~\cite{Rosenberg2015}. In the same manner, one can use CISS in transport, typically in an electrochemical setting, to employ spin in order to promote chemical reactions. For example, it was shown that when electrochemical water splitting occurs with an anode that accepts preferentially one spin owing to CISS, the process is enhanced and the formation of hydrogen peroxide is diminished~\cite{Mtangi2015,Mtangi2017,Ghosh2019,Ghosh2020}. More recently, using Hall voltage measurements it was observed that charge displacement in chiral molecules (in this case L- and D-oligopeptides) creates transient spin polarization~\cite{Kumar2017}, which in turns imparts an enantio-selective inter-molecular interaction even without electron injection. This transient spin-polarization is then also expected to affect properties at a ferromagnet/chiral molecule interface via spin-exchange interactions. The most dramatic demonstration of this principle so far was the separation of enantiomers by their interaction with
a magnetic substrate, shown in Fig.\ \ref{fig:enantio}~\cite{Banerjee2018}. The same idea, with appropriate experimental modifications, was then used for enantioselective crystallization of amino acids~\cite{Tassinari2019}. We also note that an inverse phenomenon, namely magnetization reversal in a thin-film ferromagnet solely by chemisorption of a chiral
molecular monolayer, has also been reported~\cite{BenDor2017}, with the effect possibly 
persisting for extended periods of time (hours)~\cite{Meirzada2021}.

To summarize, this short survey of experimental work shows that CISS is an important fundamental effect, with many manifestations and various practical consequences in a number of areas, from spintronic devices to chemical reactions. There is therefore great merit in theoretical understanding of CISS origins.

%% file: Theory.tex
\section{Status of Theoretical CISS studies} 
Inspired by the early experiments on CISS in transmission, theoretical studies have focused initially on scattering theory of photoelectrons off helical potentials. This was followed by a theoretical and computational focus on CISS in electron transport through helical wires, which constitutes the bulk of theory reported so far. Theory related to the more recent CISS in chemical reactions has been much more limited and is only now starting to emerge. 
We now provide a critical discussion of these efforts.

\subsection{Theory of CISS in transmission} 

Historically, scattering off asymmetric potentials is a theoretical topic 
almost as old as quantum mechanics itself. Of special importance has been the scattering of light, where Ref.\ \onlinecite{Condon1937} is an early example. 
In particular, motivated by the need to provide a theoretical underpinning for the detection of chemical helicity with polarized light, scattering of  light by {\it helical} potential shapes has been a longstanding topic, especially in theoretical chemistry and biophysics~\cite{Schmidt1970,Bustamante1980}, but also in the electromagnetic theory of chiral materials~\cite{Psilopoulos2005}. The scattering of electrons off helical - or chiral - obstacles appears to have received much less attention but more recently has been motivated by CISS experiments. 

An analytical scattering theory, which includes the effect of SOC, has been developed for electrons traversing a helical molecule~\cite{Yeganeh2009} or a self-assembled monolayer (SAM) thereof~\cite{Medina2012,Eremko2013,Varela2014}, or even a chiral molecule that is not necessarily helical~\cite{Ghazaryan2020}. Qualitatively, SOC is indeed found to induce spin-polarization. 
We note that dissipation can also polarize angular momenta. 
This can be illustrated within a classical scattering model, where the role of spin is played by classical angular momentum and SOC is replaced by friction~\cite{Hod2020}), leading us to speculate that a similar mechanism could also contribute to CISS. Unfortunately, for realistic model parameters, in particular for the SOC, the resulting polarization is too small~\cite{Gersten2013}, much smaller in magnitude than that found in some of the above-discussed experiments. 

\begin{figure}
\includegraphics[width=7cm]{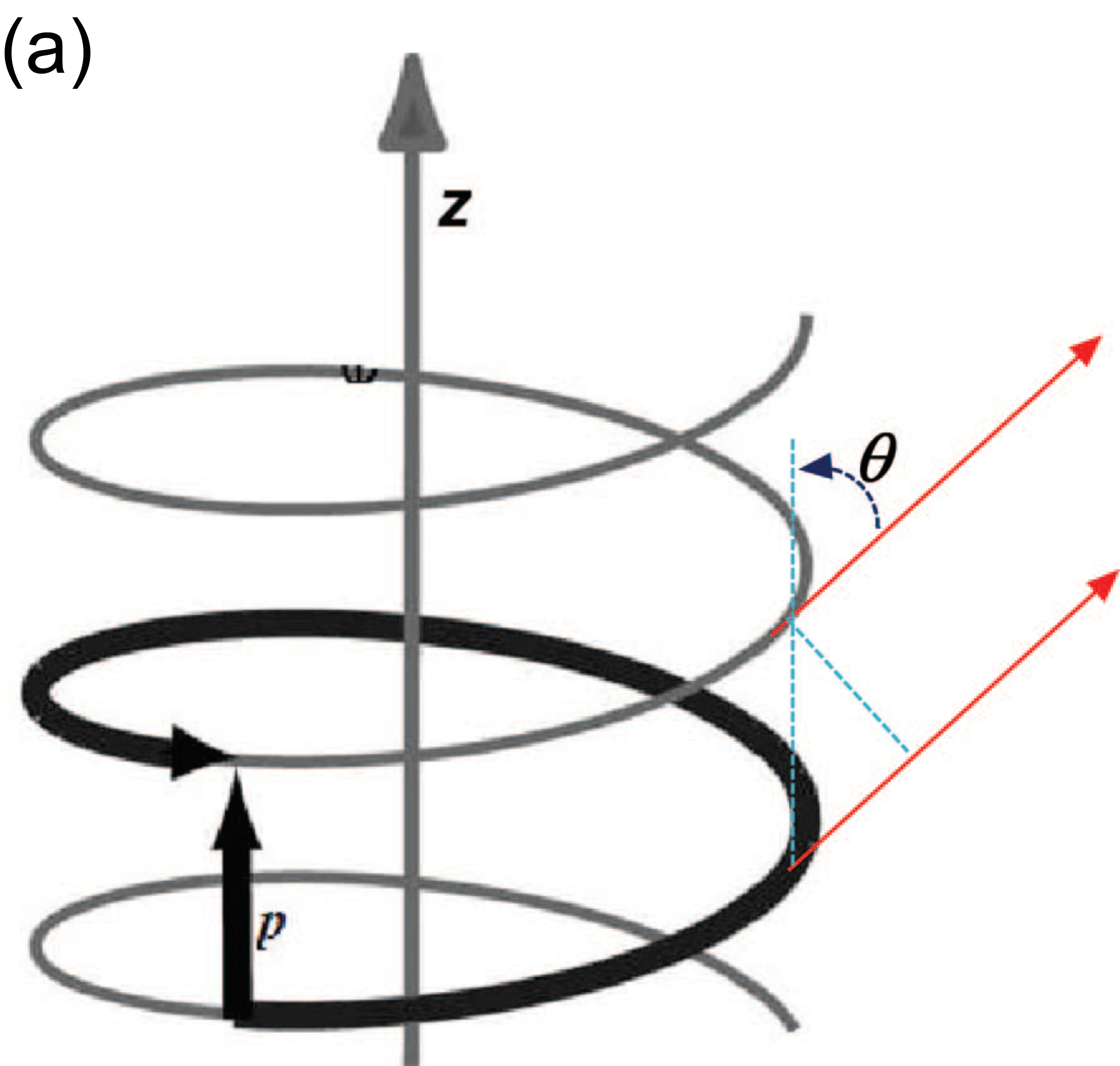}
\includegraphics[width=8cm]{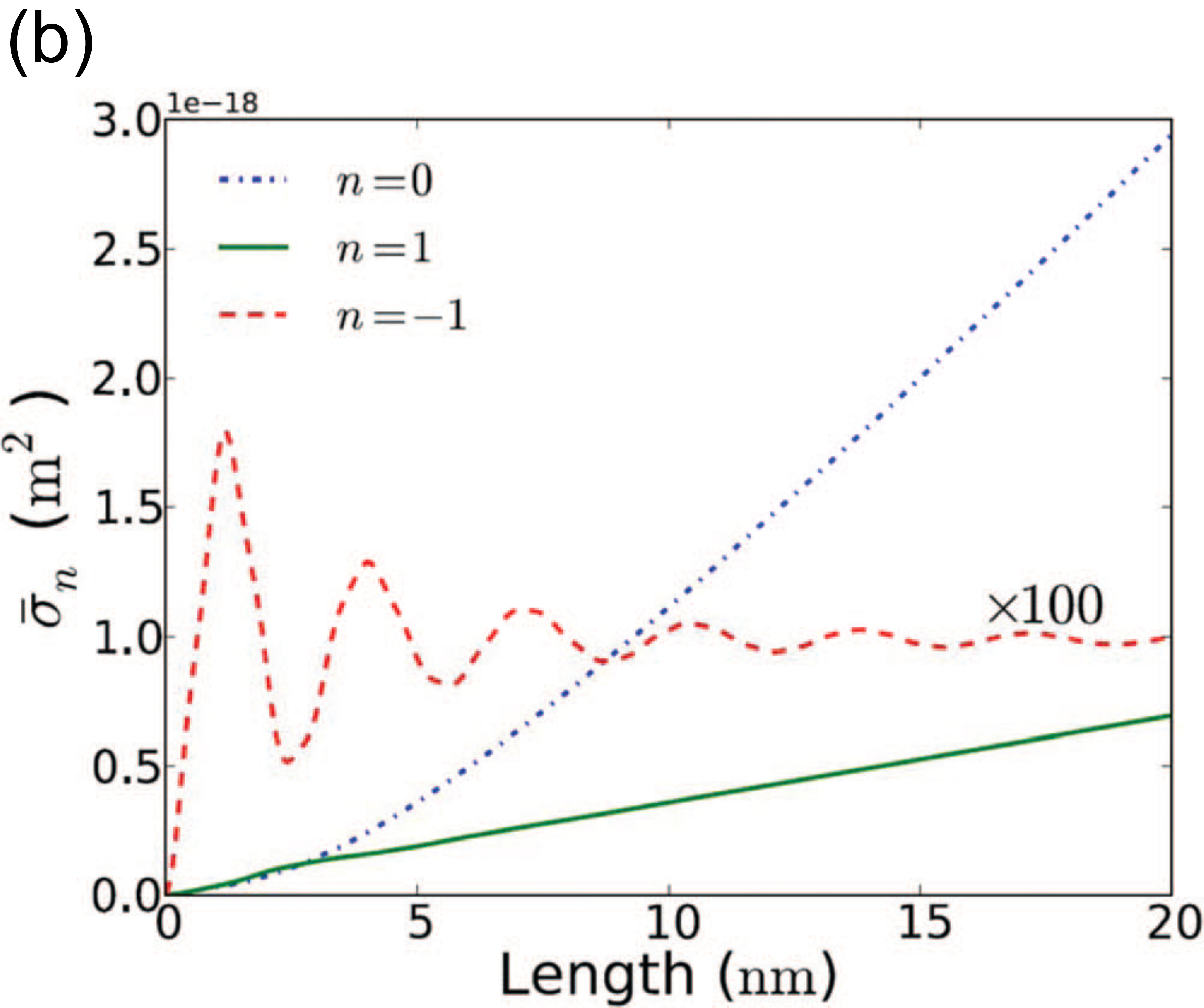}
\caption{(a) Sketch of a helical scattering potential (black), indicating the pitch (p) and the polar scattering angle, $\theta$. (b) Dependence of the (normalized) scattering cross-section 
on the angular momentum of the incident particle, for wires of increasing length $L$. Depending on  clockwise or counter-clockwise entry ($n\pm 1$), the scattering cross-section differs by two orders of magnitude. Potential parameters have been adjusted to the case of DNA; impinging energy is 0.5 eV.  
Taken from Ref.\ \protect\onlinecite{Gersten2013}, used with permission. 
}
\label{fig:scattering}
\end{figure}
In light of this discrepancy, Gersten {\it et al.}~have extended the scattering 
approach~\cite{Gersten2013} to include the SOC of the substrate that supports the SAM.
This introduces a concept of ``induced'' spin filtering, which expands the notion of ``current transfer'', developed earlier by Skourtis {\it et al.}~\cite{Skourtis2008}. The main idea is that an electron migrating through an obstacle retains a ``memory'' of its initial momentum. 
In the context of spin-filtering, this idea is used for angular momentum, 
implying that the helical molecule is ``filtering'' angular momentum. 
This means that spin-filtering occurs if the angular momentum 
of the impinging electron was at least partly spin-selected 
to begin with, as is indeed the case with a substrate possessing strong spin-orbit coupling - see Fig.\ \ref{fig:scattering}. 
While the authors expressed a hope that their theory could at least roughly 
account for the magnitude of the experimental observations~\cite{Gersten2013}, 
there appears to be no consensus concerning this claim~\cite{Matityahu2017}. Experimentally, CISS has been observed using substrates with negligible SOC.\cite{Mishra2013,Kettner2016,Kettner2018} Therefore substrate SOC may indeed be an important contribution where it exists, but cannot explain the whole effect.

\subsection{Theory of CISS in transport}

As discussed above, many \ciss-related transport phenomena have been reported experimentally, often through the measurement of  spin-resolved current-voltage 
(I-V) characteristics of electrons flowing through a chiral medium, as for example in Fig.\ \ref{fig:transport} above. It is therefore only natural that much of the theoretical effort has been focused in the same direction. 

Before considering any specifics of the these efforts, we consider what basic aspects distinguish scattering and transport experiments on a conceptual level. Taking a rough perspective, the passage of electrons from a source to a drain appears similar for bound and unbound particles: in both cases electrons migrate through a region with obstacles and therefore conventional scattering terminology applies in either situation, emphasizing a notion of similarity. However, significant conceptual differences enter upon considering that: (i) The passage of bound particles through a thin wire is quasi-one-dimensional, while scattering of an unbound particle is intrinsically three dimensional; (ii) The bound particle experiences the properties of the underlying material much more strongly than the unbound particle. For example, the electron dispersion relation will, in general, no longer be parabolic and interactions with other degrees of freedom (e.g., inelastic "multi-scattering processes") tend to be much stronger.

We begin by recalling that general conditions under which a two-terminal device can be expected to exhibit spin filtering, even in principle, have been worked out in the field of spintronics. This analysis includes CISS as a special case thereof and therefore provides an important framework for our discussion. 

We emphasize two basic facts arising from it:
(i) It is well known that in single channel
wires SOC can be 
removed by a gauge transformation~\cite{Meyer2002}. Mathematically, this is because SOC can be written as an SU(2) gauge field. Intuitively, this means that due to
the absence of loops (and magnetic fields) the spin rotates in one-to-one correspondence with a position-space shift. 
Because the spin degree of freedom can be gauged out, a spin filtering functionality based solely on SOC 
is not expected~\cite{Meyer2002}. A particular consequence of this is that spin-filtering owing to SOC is not expected to arise in tight-binding models of single-stranded DNA that afford only a single orbital per site. Therefore, Refs.\ \onlinecite{Guo2012, Gutierrez2013, Matityahu2016} have emphasized the importance of two channels for the observation of {\ciss}.

(ii) Even in two-terminal molecular junctions supporting several channels, spin-selectivity is still suppressed in the linear regime because of  
time-reversal symmetry, as emphasized in Ref.\ \onlinecite{Yang2019} (and see Refs.\ \onlinecite{Naaman2020c,Yang2021c} for additional discussion). The proof of this claim uses an Onsager-type argument:  
Consider a two-terminal device with a non-magnetic source. Let the drain exhibit magnetism, $\bM$, so that it can act as a spin-analyzer, but there is no magnetic field otherwise. 
The argument proceeds via {\it reductio ad absurdum:}
Assume that the device could act as a spin-filter. 
Then the conductance would be sensitive to the direction 
of $\bM$ in the analyzer; 
in particular $G(\bM) \neq G(-\bM)$. However, an Onsager relation 
protected by time-reversal invariance implies $G(\bM)=G(-\bM)$, 
leading to a contradiction and thereby proving the claim. 
We also note that a special case of both of the above-statements, for non-interacting single-channel wires in the linear regime, is known as the "single-channel no-go-theorem"~\cite{Kiselev2005,Bardarson2008}.

Below we survey many studies that have reported chirality-selective spin transport in models of single-channel wires. In many (but not all~\cite{Guo2012}) cases, violation of these restrictions is reported. A detailed analysis of each individual theoretical/computational model is beyond the scope of this overview. However, it is of utmost importance to assess model predictions against the above general restrictions~\cite{Yang2019,Entin2021}. In some cases, violations may be rationalized in terms of the computed quantities or the fundamental model assumptions (some examples are given below), while in others this may reveal mistakes in the analysis. 

A considerable number of studies based their analysis on transmission calculations for tight-binding models. 
\cite{Guo2012,Guo2012b,Gutierrez2012,Gutierrez2013,Guo2014,Guo2014b,Varela2016,Geyer2019,Sierra2020} In an early study, Gutierrez {\it et al.}~ \cite{Gutierrez2012} considered numerically 
a single-channel tight-binding 
model with nearest neighbor hopping and \soc.
They reported a very large spin polarization near the band edges, reaching up to 100\%. 
The degree of spin polarization obviously depends on the tight-binding model parameters.
The authors motivated their choice based on DNA, with 
the hopping parameter reported to be in the range of 20--40 meV and with chirality entering the model indirectly via its feedback into the {\soc}. 
To determine the latter, a heuristic argument 
was exploited, which yields for light atoms (C, B, N, O) 
typical (about 2 meV) values for coupling strength. 
Indeed, scales of meV can be reached with light elements, 
{\it e.g.}, when promoting a carbon atom in graphene from sp$^2$- to sp$^3$-hybridization~\cite{Gmitra2009}. 
However, as compared to this promotion, the chirality-induced symmetry breaking 
should be weaker by a geometric factor that incorporates the helical parameters of pitch and diameter. 

Clearly, further insight into a quantitative estimate of the effect can come from first principles studies that do not utilize model parameters. However, these are challenging because full helicity usually implies large molecules that need to be treated at a level of theory that includes proper relativistic corrections to the electronic structure. 

Facing these difficulties, 
Maslyuk {\it et al.}~\cite{Maslyuk2018} have attempted a first principles transport calculation for a DNA molecule. In their calculations, the {\soc} has been implemented employing pseudopotentials and the 
zero order regular approximation (ZORA)~\cite{vanLenthe1994}.
Spin-filtering of an $\alpha-$helix was compared to that of a $\beta-$strand, the latter corresponding to an enforced linear geometry. The authors found that spin-filtering was stronger in the helical conformation, as compared to the linear one, which is expected due to the symmetry reduction. 
Quantitatively, however, the observed polarization was an order of magnitude below the experimental reports. 
This observation has been shared by Rebergen and Thijssen~\cite{Rebergen2019}: At a qualitative level, the existence of {\ciss} is confirmed, but quantitatively the native {\soc} in the molecules considered appears to be too small, possibly by an order of magnitude, in order to account for the experimental results. Thus, an additional and important challenge is the same quantitative issue faced by the theory of CISS in transmission. In light of this difficulty, and inspired by the early work of Gersten {\it et al.}~\cite{Gersten2013}, Liu {\it et al.}
~\cite{Liu2021} proposed an orbital-polarization model with SOC from the electrode to interpret CISS in transport. In their model, by going through the chiral molecule electrons become orbital-polarized (an effect also obtained in model calculations~\cite{Michaeli2019}) and the orbital polarization is converted to spin polarization by the SOC in the electrodes. This leads, in the non-linear regime, to  unidirectional magnetoresistance, rationalizing CISS to some extent. 

Quantitative issues notwithstanding, the vexing problem of the microscopic origins of \ciss-type phenomena 
in transport has motivated many authors to investigate situations that definitively avoid the Onsager-based no-go theorem. In a recent example, Utsumi {\it et al.} calculated time-reversal symmetric charge and spin
transport through a molecule comprising two-orbital channels and connected to two leads. They demonstrated that spin-resolved currents are generated when spin-flip processes are accompanied by a 
flip of the orbital channels.\cite{Utsumi2020} 

\begin{figure}[ht!]
\centering
\includegraphics[width=7.5cm]{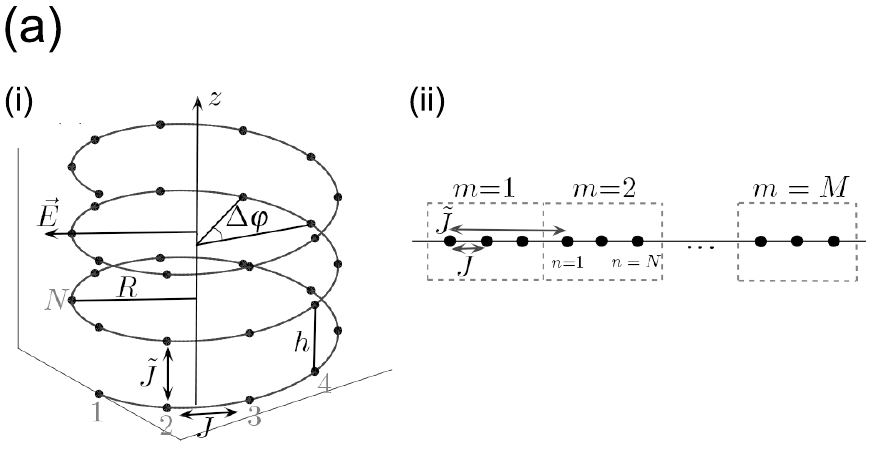}
\includegraphics[width=8cm]{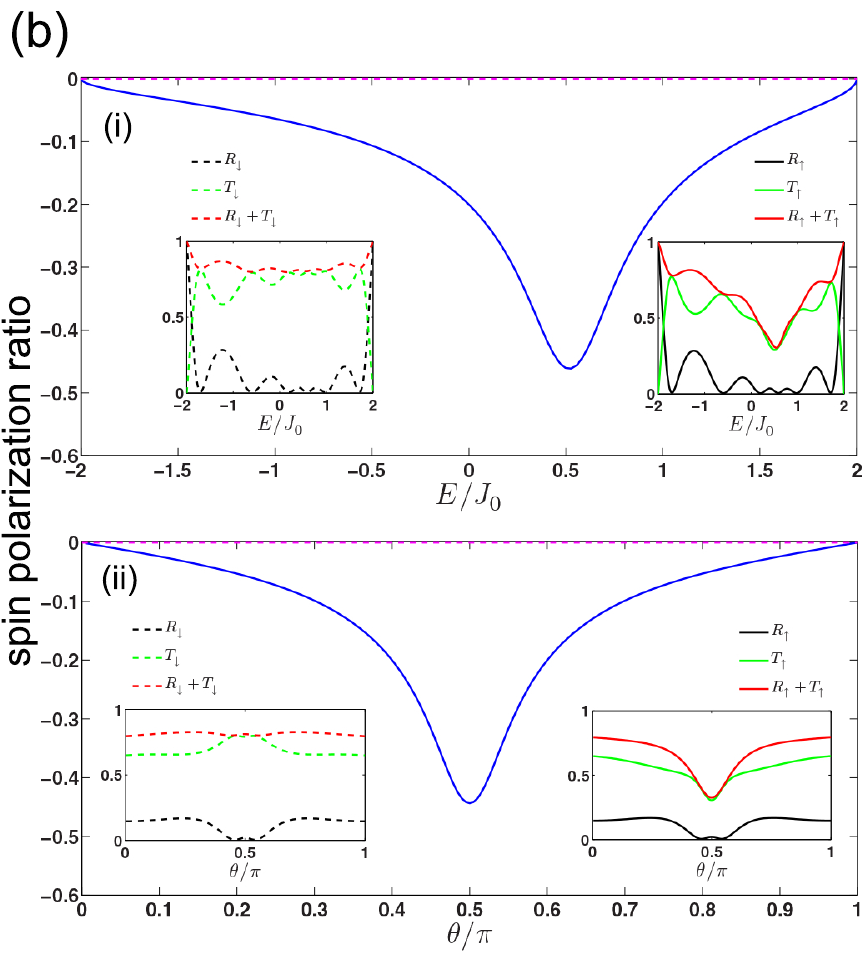}
\caption{(a)
Tight binding model of a single helical molecule with radius $R$, pitch $h$, and twist angle $\varphi$. Electrons can hop between
adjacent sites along the helix with hopping amplitude $J$ or vertically
to the $N^\mathrm{th}$ neighbor with hopping amplitude $\tilde{J}$. Spin-orbit interaction is assumed to act only between nearest neighbor sites. (i) Schematic view of the
helical molecule. (ii) Mapping of the model onto a one-dimensional chain of $M$ unit cells, each containing $N$ sites. (b) Spin polarization (solid blue) in a chain of $M$=6 unit cells (other parameters are $N$=2, $J$=1.5$J_0$, $\tilde{J}$=0.6$J_0$, with a lateral component $J_x$=0.2$J_0$, using a complex self-energy. (i) as a function of energy (in units of $J_0$), with $\theta$, an indicator of SOC strength, given by 0.4$\pi$.
(ii) as a function of $\theta$, with $E$=0 (center of the
band). The spin polarization vanishes (dashed magenta) if
either $J_x$=0 (unitary chain), $\tilde{J}$=0 (nearest
neighbor chain), or $\theta$= 0 (no SOC). Inset: Reflection
(black) and transmission (green) coefficients for spin up
(solid) and spin down (dashed). The red line is the sum of
these coefficients.
Taken from Ref.\ \protect\onlinecite{Matityahu2016}, used with permission.
}
\label{fig:leakage}
\end{figure}

Guo {\it et al.}~\cite{Guo2012,Guo2014} (and in a different context independently also 
\onlinecite{Liu2021}) and \onlinecite{Matityahu2013} 
have suggested a ``symmetry workaround'', further explored in Refs.\ \onlinecite{Matityahu2016,Matityahu2017}. This involves a third bath that the electrons traversing the chiral molecule may couple to.
The bath gives rise, in general, to non-unitary effects such as 
`dephasing' or `leakage', such that time-reversal symmetry is effectively broken
and Onsager's theorem no longer applies and spin polarization ensues. An example is shown in Fig.\ \ref{fig:leakage}.
Technically, this effect has been modeled in these studies by introducing 
an anti-hermitian self-energy $\ci \Gamma_\text{d}$. 
This was found to bring about 
spin-selective transport in the presence of {\soc}. 
An intuitive understanding of this finding can proceed from the observation that, in the presence of {\soc} and leakage, 
evanescent waves associated with opposite spins have different decay lengths~\cite{Matityahu2016}.

Quantitatively, 
this approach typically uses \cite{Guo2012,Guo2012b,Guo2014,Guo2014b,Pan2015} 
model parameters similar to those of Ref.\ \onlinecite{Gutierrez2012}, 
so that a quantitative uncertainty carries over. Furthermore, the overall magnitude of the filtering effect thus brought about is 
very sensitive to the leakage rate $\Gamma_\text{d}$. 
While the very existence of this rate is physically well motivated, 
its magnitude is still difficult to establish in realistic terms. The rates employed in the simulations to achieve spin-polarization in the experimentally 
reported regime are usually not small, typically a few percent of the 
hopping integral. Since the resulting leakage is very significant, 
the overall effect thus achieved still appears to be too small 
to explain the main experimental features, and one would further expect significant resistance effects.


A different approach to the idea of a bath has been proposed recently by Volosniev {\it et al.}, who proposed that the {\ciss}-effect is a many-body phenomenon that arises from a bath that manifests in the carrier dynamics as friction~\cite{Volosniev2021}.
To illustrate their idea, they adopted a qualitative single-channel model, in which friction enters as an effective electric field that is proportional to the product of friction constant and momentum expectation value. The latter is non-vanishing in the presence of spin-orbit interaction and points in opposite directions for different spin orientations. By construction, the model produces a quasi-stationary state with spins pointing at opposite directions for a wire of finite length. While the main idea is transparent, the relation to the {\ciss}-effect remains uncertain for two reasons: (a) The microscopic source of the friction, and therefore the relevance of the qualitative model, remains an open issue. (b) Spin-separation is brought about by a friction-controlled dynamical process, which requires the supply of a sufficient amount of energy, the source of which remains unspecified.

A natural candidate for a physical bath that participates in the carrier dynamics are the atomic nuclei~\cite{Fransson2020,Fransson2021,Zhang2020}.
As recently pointed out~\cite{Wu2021,Bian2021}, the effect of the nuclei on electronic spin separation can be enhanced due to conical intersections, so that the effect could indeed contribute to the {\ciss}-phenomenon. 

Exploring a different approach, Dalum and Hedeg{\aa}rd~\cite{Dalum2019} considered situations in which the source feeds a current into the device, i.e. the helical molecule, with an occupation of incoming scattering states that is out of equilibrium. In fact, such a situation arises very naturally when photo-electrons traverse a helical SAM; 
it is, however, more difficult to motivate in the context of conventional conductance experiments.
Dalum and Hedegard point out, in addition, that even after a workaround has been implemented, 
one still faces the problem that {\soc} is small as compared to all other native energy scales, so a sizable polarization is not expected. To overcome this difficulty, the authors invoke degeneracies, the consequences of which they study by adopting helical polyacetylene as a paradigm system.  

An altogether different approach towards understanding \ciss-type effects 
has been undertaken by Yang {\it et al.}~\cite{Yang2019}. They  take the existence of 
\ciss-effects for granted and construct, based on this assumption, 
phenomenological models that describe typical {\ciss} measurements. 
The model is formulated in terms of different versions of transfer matrices that 
represent different elements of the measurement circuit, 
such as magnetic and non-magnetic barriers, the helical molecule, etc. 
Due to the model simplicity, analytical calculations are feasible. 
By construction, Onsager's reciprocity theorems are satisfied by the model. Therefore, 
an explicit calculation of the two-terminal conductance yields the expected negative result, i.e., no spin-filtering in a linear-response two-terminal calculation. However, a positive result is found in multi-terminal calculations. 

A puzzling aspect of Ref.\ \onlinecite{Yang2019} is that on the one hand the existence of {\ciss} is deduced from experiments, which have been performed in two-point geometries, while on the other hand it is found that two-point conductance 
measurements will not show the {\ciss} effect. Routes escaping this dilemma are 
proposed in Ref.\ \onlinecite{Yang2020}. Specifically, two natural routes are discussed : (i) The reciprocity theorem makes no statement about non-linear effects; therefore, traces of {\ciss} can exist, and have been identified in Ref.\  \onlinecite{Yang2020}, also in two-terminal measurements if they are operated in the non-linear regime of bias voltages. (ii)
Also, the reciprocity theorem does apply if time reversal invariance is broken. Hence, one might expect that aligning two helical molecules in series with a small resistor in between will yield different results, depending on whether the molecules have the same or opposite helicities. Along this idea another set of experiments has been proposed in Ref.\ \onlinecite{Yang2020}, which could show CISS in a two-terminal setup, here even in the linear regime.


Finally, we mention a conceptually interesting field-theoretical approach towards {\ciss} that has recently been put forward~\cite{Shitade2020}. It considers the Dirac equation in a  one-dimensional curved space-time. In this framework, the usual Foldy-Wouthousen transformation can be used in order to project into the non-relativistic (low kinetic energy) sector. The authors show that as a result of curvature, the kinetic energy acquires an SU(2)-gauge field, which plays a role analogous to the {\soc} in single-channel wires. Embarking on this observation, the authors apply their theory to transport in helical molecules, where the (quasi-)one-dimensional molecule is interpreted as a physical realization of a curved space-time for the traversing electron. The basic idea of this application is conceptually appealing, but the space-time considered in Ref.\ \onlinecite{Shitade2020} carries only a single parameter, the curvature $\kappa$ of the helical path. As a consequence, the theory does not describe the three-dimensional nature of experimental observations, which manifests as an emergence of molecular properties when gradually adding atom by atom. Furthermore, the results depend on the order of limits (first dimensional reduction, then non-relativistic limit), a point also emphasized by Geyer {\it et al.}~\cite{Geyer2020}.

\subsection{Theory of CISS in chemical reactions}

As explained above, CISS in chemical reactions is a very new field, even experimentally. Accordingly, theory is still limited. First principles calculations have repeatedly shown that if one assumes that chirality indeed begets spin polarization, then the latter can explain chemical enantio-selectivity~\cite{Kumar2017,Banerjee2018}. Recent first principles calculations for a chiral monolayer on a magnetic substrate provided first indications of an emergent electronic structure and emphasized the role of exchange interactions~\cite{Dianat2020}. However, the degree to which the emergent structure is affected by the choice of density functional~\cite{Eckshtain2016} was not investigated. In addition, chiral symmetry breaking was enforced ``manually'' via introduction of an external Ni atom, so that the effect of the intrinsic chirality remains to be clarified. To the best of our knowledge, a more complete theoretical framework that describes how CISS emerges in such scenarios has yet to be provided.

%% file: Conclusions_Outlook.tex
\section{Conclusions and Outlook} 

In this overview, we have surveyed the three main types of CISS - in transmission, transport, and chemical reactions. For each, we critically overviewed existing theoretical approaches, while emphasizing advantages and disadvantages with respect to qualitative and quantitative agreement with known experimental results.

At present, a unifying scheme that would allow one to interpret all experiments in terms of only a single microscopic effect – the ``CISS effect'' - has not yet been identified. While such a framework cannot be ruled out, chirality-induced spin selectivity may perhaps be thought of as a set of phenomena that have a unifying scheme only in the sense that they all derive from the interplay of spin-orbit interaction and chirality. For example, it has been suggested theoretically that spin-orbit interaction 
leads to non-conservation of spin currents in a two-terminal junction and consequently to
a mechanical torque~\cite{Sasao2019}, which is a different experimental observable than the ones surveyed above. Such a CISS-induced torque has indeed been suggested as an explanation for a recent experiment 
demonstrating the use of a chiral molecule as a molecular motor~\cite{Wulfhekel2021}. 

Correspondingly, a large gap remains between the experimental observations and the quantitative estimates from theory. Further experiments are required for guiding the theory and for limiting
the possible interpretations for this potentially very important phenomenon. Further exploration of recent suggestions of finite temperature effects, which go beyond pure electronic ones, is also of interest.

Finally, theoretical studies have focused on steady-state  transport, but very little theory addresses the growing number of experiments that report transient CISS phenomena (e.g.~\cite{Kumar2017}). Likewise, the hypothesized role that such transient phenomena may play in CISS in chemical reactions~\cite{Banerjee2018} has not been sufficiently explored yet. 

We believe that consolidating the field, in the sense of bringing theory and experiment much closer, with the goal of achieving a detailed microscopic understanding of CISS, is an ongoing challenge that provides many research opportunities. To this end, one can perhaps concentrate on model systems, where CISS  can be studied in great detail in both experiment and theory. 